 \documentclass[final,5p,times,twocolumn]{elsarticle}

\usepackage{amssymb}
\usepackage{lipsum}
\usepackage{xcolor}
\usepackage{siunitx}
\usepackage{bm}
\usepackage{comment}
\usepackage{amsmath,amssymb,amsfonts}%
\usepackage{siunitx}
\usepackage{booktabs}
\usepackage{array}
\usepackage{soul}
\usepackage{wrapfig}
\usepackage{hyperref}
\hypersetup{
    hidelinks
}

\journal{Current Opinion in Colloid \& Interface Science}

\begin{document}

\begin{frontmatter}
\title{\textcolor{black}{Recent Modeling Advances in Dense Suspension Rheology}}
\author[one]{Orhun Ayar}
\author[two]{Bhargav Sriram Siddani}
\author[two]{Ishan Srivastava}
\author[one]{Abhinendra Singh}
\affiliation[one]{organization={Department of Macromolecular Science and Engineering, Case Western Reserve University},
            addressline={Cleveland, OH 44106}, 
            country={USA}}
\affiliation[two]{organization={Center for Computational Sciences and Engineering, Lawrence Berkeley National Laboratory},
            addressline={ 1 Cyclotron Rd, Berkeley, CA 94720}, 
            country={USA}}
\begin{abstract}
Dense suspensions of particles dispersed in liquids are central to industrial and geophysical processes and serve as model systems for out-of-equilibrium soft matter. At high particle concentrations, they exhibit stress-dependent rheology, including discontinuous shear thickening and shear jamming, arising from frictional contacts. Nonlinear physics arises from the interplay among direct contacts, interfacial chemistry, and fluid-mediated hydrodynamics. The relative importance of these mechanisms depends on the particle properties and flow conditions, making predictive modeling inherently multi-scale and, therefore, computationally challenging. Recent advances in computational methods have transformed our ability to simulate the physics of dense suspensions across different scales. 
\textcolor{black}{In this Perspective, we primarily focus on} state-of-the-art simulation frameworks that integrate the mechanics of dry granular materials, mediated by contact friction, with suspension hydrodynamics to provide predictive models of dense suspension rheology. We highlight recent computational developments for simulating \textcolor{black}{dense, predominantly non-Brownian suspensions in the viscous limit} at varying levels of fidelity, ranging from particle-resolved to continuum models, as well as models that investigate their mesoscale organization during flow. Together, these approaches reveal a hierarchical structure in which microscale constraints give rise to mesoscale frictional networks that ultimately govern the macroscopic flow. \textcolor{black}{
While our emphasis is on dense suspensions governed by near-field hydrodynamics and frictional interactions, we also briefly discuss complementary grid-based methods that can account for complex geometries and particle properties, and also enable continuum-scale descriptions.}
\end{abstract}



\begin{keyword} dense suspensions \sep colloids \sep simulations \sep friction \sep hydrodynamics \sep network science
\end{keyword}
\end{frontmatter}

\section{Introduction}
\label{introduction}
\begin{figure*}
	\centering 
	\includegraphics[width=0.9\textwidth,trim={0 0 0 0},clip]{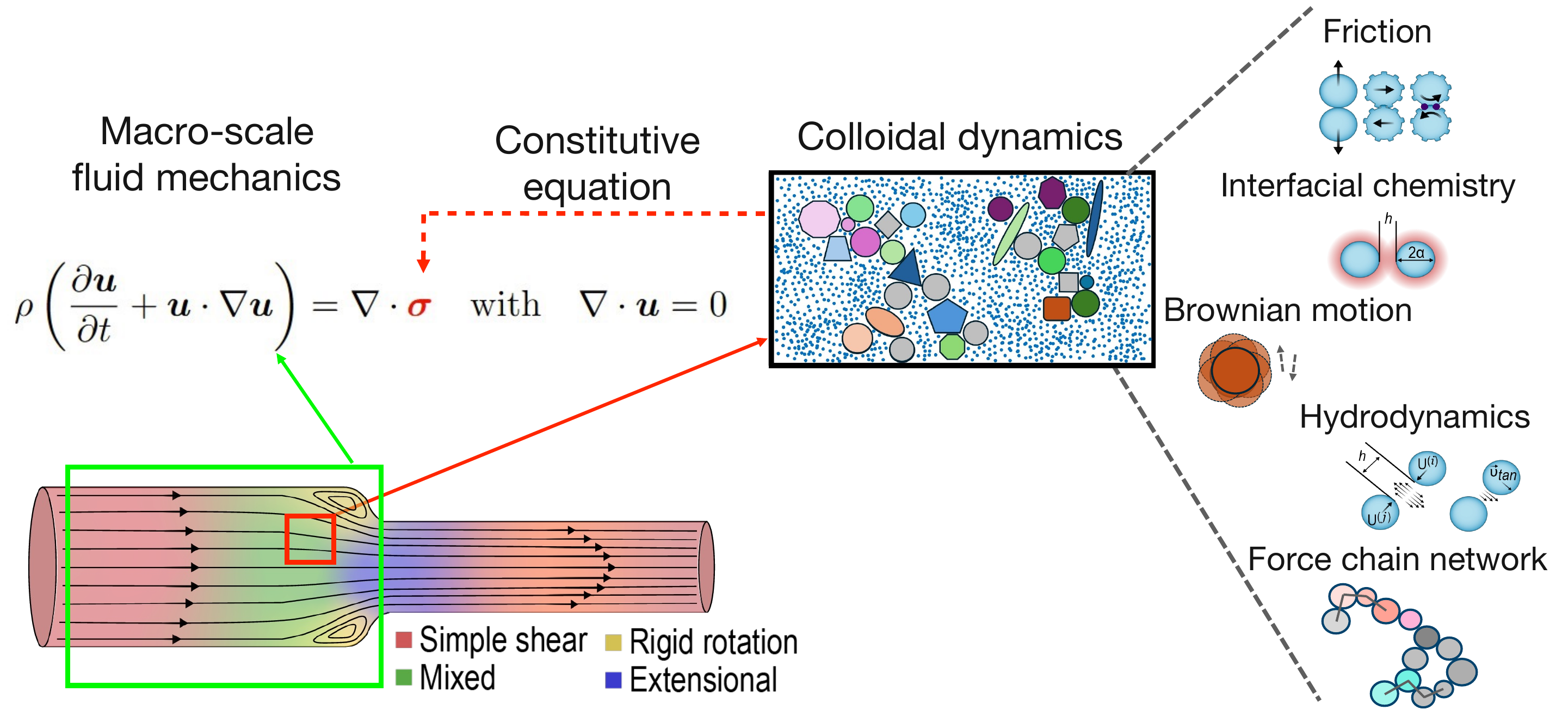}	
	\caption{\textbf{Multi-scale nature of suspension flow: The channel flow through a contraction offering a paradigmatic example} \emph{Top Left}: Role of constitutive equation in macro-scale fluid mechanics governed by constitutive equation. The constitutive equation is governed by colloidal dynamics, which is in turn governed by forces and constraints at play: attraction-repulsion interactions, near-field hydrodynamics, Brownian motion, rolling and sliding friction constraints, and force chain networks.
    } 
	\label{fig_schematic}%
\end{figure*}

Dense suspensions are prototypical complex fluids whose flow behavior is challenging to predict yet is central to industrial, geotechnical, and biological processes such as cement transport, mudflow, and blood flow~\cite{Morris_2020, Jerolmack_2019, hodgson2022granulation}. In the quiescent state, dense and confined suspensions, the particle motion exhibits pronounced hindrance and anomalous diffusion. Under shear, they display nonlinear responses including large normal stresses, shear thickening, shear jamming, and yielding \cite{Ness_2022, clavaud2025quick}. Because stress is shared between the fluid and particulate phases and is highly sensitive to particle concentration and evolving microstructure, predictive modeling remains fundamentally challenging. Although these behaviors have long been observed experimentally, their quantitative description requires the consistent integration of hydrodynamic interactions, contact mechanics, and multi-scale modeling strategies (Fig.~\ref{fig_schematic}). 

\begin{figure}
\vspace{-2.3em}
	\centering 
	\includegraphics[width=0.35\textwidth]{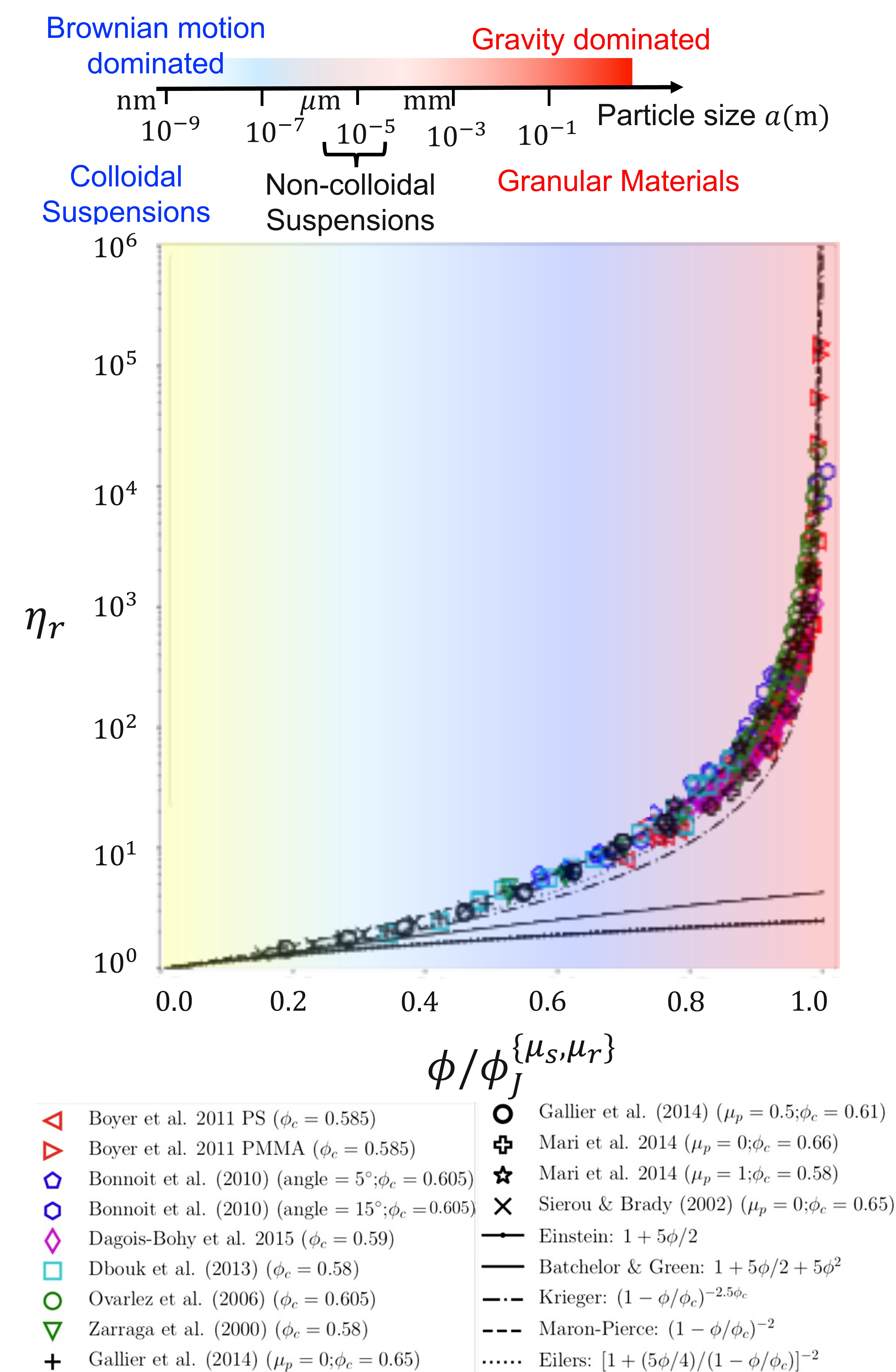}	
	\caption{\textbf{Definitions of flow regimes} \emph{Top: Regimes defined based on particle size $a$}: At the smallest length scale, Brownian motion and hydrodynamics dominate, while gravity and friction govern the physics in the granular regime; both friction and hydrodynamics are crucial for the non-colloidal suspensions, and particle inertia can be ignored. \emph{Bottom Flow regimes based on scaled particle concentration in rheology of colloidal and non-colloidal suspensions}: From dilute (yellow shading) to non-dilute (blue) and dense (red) limits based on the relative importance of interactions; full long-ranged hydrodynamic interactions are crucial in dilute limit; near-field lubrication interaction are important in semi-dilute limit; frictional contact interactions dominate the physics in dense limit. Bottom panel is adapted from Reference~\cite{guazzelli_2018}.} 
	\label{fig_regimes}%
\end{figure}

For classification purposes, we distinguish the suspension classes based on particle volume fraction $\phi$ and characteristic particle size $a$ (Fig.~\ref{fig_regimes}). In the dilute regime, long-range hydrodynamic interactions dominate. In the semi-dilute regime, near-field lubrication interactions become important, although persistent contacts remain infrequent. In the dense regime, which is the focus of this Perspective, short-range interactions and near-contact physics govern stress transmission, as the mean interparticle separations become much smaller than the particle size~\cite{Ness_2022,clavaud2025quick}. Suspensions can also be distinguished by particle size $a$, where dimensionless numbers are helpful: the Reynolds number, $Re = \rho_f a^2 \dot{\gamma}/\eta_0$, and the Stokes number, $St = \rho_p a^2 \dot{\gamma}/\eta_0$, which compare fluid and particle inertia, respectively, to viscous stresses. We focus on the viscous limit with $Re \ll 1$ and $St \ll 1$. The P\'eclet number, $Pe = 6\pi \eta_0 a^3 \dot{\gamma}/k_B T$, quantifies the relative importance of shear to Brownian motion:  $Pe \lesssim 1$ corresponds to colloidal suspensions, while $Pe \gg 1$ characterizes athermal non-colloidal systems. Granular suspensions may also involve gravitational and inertial effects.

The experimental characterization in the dense limit is intrinsically challenging. Wall slip, shear-induced migration, free-surface instabilities, and sample ejection complicate the measurements and obscure the internal stress state~\cite{Morris_2020}. Simultaneous access to shear stress, particle pressure, and normal stress differences remains limited under strongly non-Newtonian conditions, and resolving particle-scale dynamics and contact organization in situ is challenging.

Therefore, computational simulations provide an essential complementary approach to experimental studies. By resolving particle motion under controlled conditions, simulations connect bulk rheology to particle-scale dynamics and evolving microstructure while enabling systematic variations in the interaction laws and loading conditions. In the dense regime, where lubrication, contact, and frictional interactions compete, particle-resolved computational frameworks have become indispensable for linking microscopic physics to emergent rheology.

In this Perspective, we synthesize recent advances in modeling approaches for suspension rheology, 
\textcolor{black}{with a primary focus on dense, predominantly non-Brownian suspensions in the viscous (low-Reynolds-number) limit}. 
\textcolor{black}{In this regime, nearly similar-sized particles dispersed in a Newtonian solvent interact through hydrodynamic, conservative, and contact forces, which together govern the rheological response.} 
As a Perspective, our goal is \textcolor{black}{not to provide an exhaustive review of all suspension classes or modeling approaches}. 
\textcolor{black}{Instead, we emphasize a physically grounded framework in which microscale interactions give rise to constraints, constraints organize mesoscale frictional contact networks, and these networks govern macroscopic flow behavior.} 
\textcolor{black}{Where appropriate, we briefly discuss extensions beyond this regime, including effects of particle inertia, non-spherical particles, confinement, and solvent complexity, as well as continuum--particle coupling approaches, to provide broader context and highlight emerging directions.}

\section{A Rapid Overview of Methods}
Modeling the hydrodynamic interactions between suspended particles and coupling them consistently with other relevant interactions, such as colloidal, contact, and conservative forces, is often the most challenging aspect of suspension simulations. This difficulty is further exemplified by the long-range nature of solvent-mediated interactions. In general, simulation approaches can be classified into implicit and explicit solvent methods, depending on the representation of the fluid phase. 

\paragraph{Implicit solvent models}
In implicit solvent models, the effect of the surrounding fluid is incorporated through effective hydrodynamic forces acting between particles. In mesoscale systems, additional stochastic forces are included to represent the thermal fluctuations. These approaches are typically derived from a continuum description of the solvent. In the low-$Re$ regime, the linearity of the Stokes equations allows hydrodynamic forces and torques to be linearly related to the particle velocities through a resistance or mobility tensor. Therefore, much of the development of implicit solvent methods has focused on constructing accurate and computationally tractable approximations of this tensor. Analytic representations are typically limited to simple solvent properties and particles with high shape symmetry, such as spheres with moderate size dispersity~\cite{Brady_1985, morris2018lubricated}. Consequently, the applicability of implicit methods is restricted when modeling more complex suspensions encountered in practice. Additional challenges arise from the numerical treatment of the mobility tensor, particularly in schemes that require evaluation of its divergence to maintain consistency with the fluctuation-dissipation theorem. These challenges are further exacerbated in dense suspensions, where particle contacts dominate.

\paragraph{Continuum--particle coupling approaches} A related class of methods couples particle dynamics to a continuum fluid solver, using the finite-volume or lattice-Boltzmann method, in which the solvent equations are solved on a grid and momentum is exchanged between the Eulerian fluid and Lagrangian particles. {\color{black}{Immersed boundary (IB) ~\cite{Uhlmann2005,breugem2012second,Azis2019},  lattice-Boltzmann (LB)~\cite{Ladd2001,MaxeyReview2017,Aidun2010}, and fictitious domain (FD)~\cite{patankar2000new,glowinski2001fictitious} methods represent approaches where a particle's surface, and therefore volume, is adequately resolved to impose boundary conditions. The distinction among these approaches exists in the imposition of the fluid-particle boundary conditions. Unresolved approaches belonging to this class of computational methods include the \textcolor{black}{Computational Fluid Dynamics--Discrete Element Method} (CFD--DEM) methodology, which relies on closure relations for fluid-particle coupling.}} 
%
All these approaches offer flexibility in handling complex particle shapes, size distributions, and boundary conditions. However, incorporating dense suspension rheology and accurately resolving near-contact interactions remains an active area of research~\cite{biegert2017collision,burgisser2005addressing}.

\paragraph{Explicit solvent models} In explicit solvent models, the fluid is represented as a collection of interacting particles whose collective dynamics gives rise to hydrodynamic behavior. Popular examples include dissipative particle dynamics~\cite{Martys_2005,boromand2015viscosity} and multi-particle collision dynamics~\cite{Howard_2019,wani2022diffusion}.
These methods are attractive for their relative ease of implementation and ability to handle complex geometries and particle shapes. Similarly, accurately reproducing the incompressible solvent behavior and achieving realistic transport coefficients, particularly at large Schmidt numbers, remains challenging.

The selection of a particular method depends on the application, as each method class involves a trade-off between physical fidelity, computational cost, and applicability to the dense suspensions. In the remainder of this Perspective, we focus primarily on implicit-solvent, particle-based approaches that are most commonly used to probe the rheology and microstructure of dense suspensions.

\section{Simulation frameworks for suspensions}
The simulation of suspensions has traditionally been approached from a fluid-mechanics perspective, in which the particle motion is governed by hydrodynamic interactions among all particles, mediated by the viscous fluid (solvent). Although this provides a natural starting point for describing suspension mechanics, recent discoveries have shown that it becomes insufficient in the dense limit, thereby motivating the development of new physics. In this section, we first discuss the fluid-mechanical framework for suspensions, followed by the latest developments in near-contact physics that are required for the accurate modeling of highly dense suspensions. 

In any particle-based system, the particle motion is governed by the balance of hydrodynamic, contact, and other interparticle forces and torques, which can be written schematically as
\begin{equation}
    \boldsymbol{M}
    \frac{d}{dt}
    \begin{pmatrix}
    \boldsymbol{U} \\
    \boldsymbol{\Omega}
    \end{pmatrix}
    =
    \sum_{\alpha}
    \begin{pmatrix}
    \boldsymbol{F}_{\alpha} \\
    \boldsymbol{T}_{\alpha}
    \end{pmatrix},
    \label{eq:Motion}
\end{equation}
where $\boldsymbol{U}$ and $\boldsymbol{\Omega}$ are the translational and rotational velocities, $\boldsymbol{F}_{\alpha}$ and $\boldsymbol{T}_{\alpha}$ denote the forces and torques, respectively, and $\boldsymbol{M}$ is the mass and moment-of-inertia matrix. In suspensions, the bulk stress (excluding the isotropic fluid pressure) consists of a solvent contribution associated with the imposed rate-of-strain tensor $\boldsymbol{E}^{\infty}$ and a particle contribution arising from interparticle interactions, which can be expressed as
\begin{equation}
    \boldsymbol{\Sigma}
    =
    2\eta_0 \boldsymbol{E}^{\infty}
    +
    \boldsymbol{\Sigma}^p,
    \qquad
    \boldsymbol{\Sigma}^p
    =
    \boldsymbol{\Sigma}^H
    +
    \boldsymbol{\Sigma}^C
    +
    \boldsymbol{\Sigma}^{\mathrm{cons}},
\end{equation}
where the particle stress includes hydrodynamic, contact, and conservative contributions,
each obtained from volume-averaged stresslets,
$\boldsymbol{\Sigma}^{\alpha} = V^{-1}\sum \boldsymbol{S}^{\alpha}$, where \(\alpha \in \{H, C, \mathrm{cons}\}\).

From $\boldsymbol{\Sigma}$, standard rheological observables can be defined: shear stress $\sigma_{xy}=\Sigma_{12}$, particle pressure $\Pi = -(\Sigma_{11}+\Sigma_{22}+\Sigma_{33})/3$, and normal stress differences $N_1=\Sigma_{11}-\Sigma_{22}$ and $N_2=\Sigma_{22}-\Sigma_{33}$. The relative viscosity is $\eta_r=\sigma_{xy}/(\eta_0\dot{\gamma})$, where indices 1, 2, and 3 denote the flow, gradient, and vorticity directions, respectively.

\paragraph{\textit{Stokesian Dynamics}} For typical particle sizes of colloidal and non-colloidal suspensions, both liquid and particle inertia are negligible, i.e., $St=0$. In the overdamped limit, hydrodynamic interactions between particles arise from the disturbance flows generated by their motion in a viscous solvent. These interactions are long-ranged and many-body in nature, and they decay slowly with increasing interparticle separation. The Stokesian dynamics (SD) modeling framework was developed to systematically capture these effects by exploiting the linearity of the Stokes equation and relating particle velocities to hydrodynamic forces and torques via a grand resistance (or mobility) matrix~\cite{Brady_1985,Bossis_1989}. Given $St=0$, the equation of motion (Eq.~\eqref{eq:Motion}) reduces to:
\begin{equation}
0 =
\boldsymbol{R}
\begin{pmatrix}
\boldsymbol{U}-\boldsymbol{U}^{\infty} \\
\boldsymbol{\Omega}-\boldsymbol{\Omega}^{\infty}
\end{pmatrix}
+
\begin{pmatrix}
\boldsymbol{F}_{\mathrm{nH}} \\
\boldsymbol{T}_{\mathrm{nH}}
\end{pmatrix},
\label{eq:Stokes}
\end{equation}
where $\boldsymbol{X}$ denotes the particle positions, $\boldsymbol{R}$ is the hydrodynamic resistance matrix, $\boldsymbol{U}^{\infty}$ and $\boldsymbol{\Omega}^{\infty}$ denote the imposed background flow, and $\boldsymbol{F}_{\mathrm{nH}}$ and $\boldsymbol{T}_{\mathrm{nH}}$ represent all non-hydrodynamic forces and torques acting on the particles, respectively. The resistance matrix incorporates both far-field hydrodynamic interactions, which are typically represented through multipole expansions, and near-field lubrication forces obtained from asymptotic two-body solutions~\cite{Jeffrey_1984,Kim_1991}.
Stokesian Dynamics has been remarkably successful in describing the microstructure and rheology of dilute and semi-dilute suspensions, by accurately modeling key suspension physics such as shear thinning and weak shear thickening associated with the so-called ``hydro-clusters'', which are transient particle aggregates held together by dissipative lubrication forces~\cite{Bossis_1989}.

\paragraph{\textit{Computational advances within hydrodynamic frameworks}} The long-ranged nature of hydrodynamic interactions and the resistance matrices used in Stokesian Dynamics are computationally expensive to compute: it involves the calculation of the $\mathcal{O}(N^2)$ far-field mobility matrix and involves the inversion of the matrix $\mathcal{O}(N^3)$. Accelerated Stokesian Dynamics (ASD) and related Ewald-based approaches reduce the computational cost of far-field hydrodynamics to $\mathcal{O(N\log N)}$, allowing the simulation of significantly larger systems while retaining accurate near-field lubrication forces~\cite{Sierou_2001}. This is achieved by resolving long-range interactions by calculating the Ewald-summed wave-space contribution as a Fourier-transform sum and by inverting the now-sparse resistance matrix iteratively~\cite{Sierou_2001}, \textcolor{black}{which at the time enabled researchers to simulate the short-time self-diffusion coefficient with $\mathcal{O}(10^3)$ particles.} 
More recent formulations, such as Fast Stokesian Dynamics, recast the hydrodynamic problem to avoid the explicit construction and inversion of resistance matrices, thus extending the accessible system sizes~\cite{Swan2019FSD}. FSD avoids explicit inversion of the hydrodynamic operators, drastically reducing the computation to $\mathcal{O(N)}$, \textcolor{black}{where with a graphical unit enabled version of the Fast Stokesian Dynamics code simulated $\mathcal{O}(10^5)$ particle system.} 
These computational advances,\textcolor{black}{such as graphics processing units enabled Verlet list approaches \cite{Howard2016}}, have reduced computational costs and enabled large-scale simulations of suspensions within hydrodynamic-only frameworks.
%

\paragraph{\textit{Lubrication breakdown and near-contact physics}} 
For two smooth hard spheres, the normal lubrication resistance diverges as $F_{\mathrm{lub}} \sim \frac{1}{h}$, where $h$ is the surface separation~\cite{Jeffrey_1984,Kim_1991}. As such, in a purely continuum description of suspensions, such a divergence implies that perfectly smooth particles should never touch or make contact. Melrose and Ball demonstrated in their seminal paper~\cite{Melrose_1995} that this leads to unphysical behavior as particle surfaces reach separations far below molecular or asperity length scales. For example, these models predict unphysical pathological jamming at volume fractions well below the experimentally observed jamming conditions or even random close packing (RCP $\approx 0.64$)~\cite{OHern_2003}. \textcolor{black}{This observation implies lubrication breakdown, which can allow the formation of frictional contacts, especially relevant in dense conditions, as discussed next.}
%

\paragraph{\textit{Near-hard-sphere viewpoint and missing physics}} A consensus that has emerged in the last decade is that real particles are not perfectly smooth hard spheres; they exhibit surface roughness, finite compliance, and short-range surface forces that are dominant at separations much smaller than the particle radius $h \approx 10^{-3}a$~\cite{Morris_2023, Ness_2022}. The near-hard-sphere (NHS) modeling framework incorporates this by allowing lubrication interactions to be regularized at small separations, leading to contact friction upon physical contact between particles~\cite{Morris_2023}. Recent simulations have extensively demonstrated that by incorporating lubrication breakdown and contact friction, both discontinuous shear thickening and shear jamming can be successfully reproduced
\cite{Seto_2013a,Singh_2018,Singh_2020, Ness_2016, More_2020b}.

Few studies have examined whether shear thickening in dense suspensions can be captured within modified hydrodynamic frameworks without explicitly introducing new contact physics. Jamali and Brady~\cite{Jamali_2019} modeled surface roughness by resolving discrete asperities using dissipative particle dynamics, demonstrating that lubrication interactions between asperities can dominate energy dissipation at very small separations and give rise to the DST. In a related study~\cite{Wang_2020}, it was shown that enhancing the tangential divergence of lubrication resistance, using parameters related to asperity geometry, can also reproduce DST within a hydrodynamic description. 
\textcolor{black}{Although successful in reproducing DST, SJ remains a challenge for such hydrodynamic treatments, given that fluid-mediated forces vanish in the limit of a zero shear rate.}
%
%

Taken together, these observations motivate the near-hard-sphere (NHS) framework, in which particles interact through lubrication forces together with short-ranged conservative and contact interactions that become relevant at small surface separations, thus providing a minimal description of dense suspension rheology~\cite{Morris_2023, clavaud2025quick, Singh_2020}. 

\paragraph{\textit{Integrating hydrodynamics with contact friction in dense suspensions}} 
Following Melrose and Ball~\cite{Melrose_1995}, hydrodynamic interactions are typically restricted to single-particle Stokes drag and pairwise lubrication forces, which dominate dissipation in the crowded environments. Once the particles approach a small cutoff separation $h \le 10^{-3}a$, a distance that mimics the small asperities on the surface of the particle, the lubrication forces are regularized ($F_{\mathrm{lub}} \sim \frac{1}{h+\Delta}$), allowing frictional contacts between the particles. 
Thus, Stokes flow (Eq.~\eqref{eq:Stokes}) is
\begin{equation}
\vec{0}
=
\vec{F}_{\mathrm{H}}(\vec{X},\vec{U})
+
\vec{F}_{\mathrm{C}}(\vec{X})
+
\vec{F}_{\mathrm{cons}}(\vec{X}),
\label{eq:LFDEM}
\end{equation}
where $\vec{F}_{\mathrm{H}}$, $\vec{F}_{\mathrm{C}}$, and $\vec{F}_{\mathrm{cons}}$ denote the hydrodynamic (including lubrication), contact, and conservative forces, respectively.

Contacts are modeled using granular physics, following the discrete element method (DEM) originally introduced by Cundall and Strack~\cite{Cundall_1979}. When two particles come into contact, i.e., overlap $\delta^{(i,j)}\ge 0$, a normal contact force $\boldsymbol{F}_{C,\mathrm{n}}^{(i,j)} = k_n \delta^{(i,j)} \boldsymbol{n}_{ij}$ is introduced along their line of centers. Here, $k_n$ is the normal spring stiffness and $\boldsymbol{n}_{ij}$ is the unit vector connecting the particle centers.
Tangential contact forces are modeled through an elastic spring that accumulates tangential displacement as $\boldsymbol{F}_{C,\mathrm{t}}^{(i,j)} = k_t \boldsymbol{\xi}_{ij}$ with $k_t$ and $\boldsymbol{\xi}_{ij}$ being tangential spring stiffness and the spring stretch, respectively. These forces are subjected to the Coulomb friction criterion $|\boldsymbol{F}_{C,\mathrm{t}}^{(i,j)}| \le \mu |\boldsymbol{F}_{C,\mathrm{n}}^{(i,j)}|$. In real-life suspensions with adhesive surface chemistries, the particles are often rough with asperities and have faceted shapes, which leads to additional friction (constraints on relative motion), i.e., rolling and twisting modes~\cite{Singh_2020, Santos_2020}. In the dry granular literature, it has long been recognized that contacts between non-spherical shapes can be modeled using spherically symmetric shapes with additional constraints, and thus torques~\cite{Estrada_2011}.


This framework can be extended to regimes with small but finite particle inertia, allowing a unified treatment of viscous and inertial contributions within the same numerical approach~\cite{Ness_2016, Ness_2023}. In this case, inertia enters as an additional time scale without altering the underlying role of the near-contact interactions and frictional contacts. Although the simplest implementations assume a constant $\mu$, recent studies have shown that $\mu$ can depend on the history of deformation or contact load, reflecting the elastic and plastic deformation of the surface asperities~\cite{Gallier_2014, More_2021,ruiz2023tribological}. Incorporating such deformation-dependent friction laws provides a more realistic description of particle contacts, particularly at high stresses, \textcolor{black}{which remains an active research topic.}

Together, these developments establish particle-based simulations that explicitly resolve near-contact interactions and frictional forces as the minimal framework for understanding the physics of dense suspension flow beyond purely hydrodynamic descriptions.

\begin{figure*}[!hbtp]
	\centering 
	\includegraphics[width=0.95\textwidth]{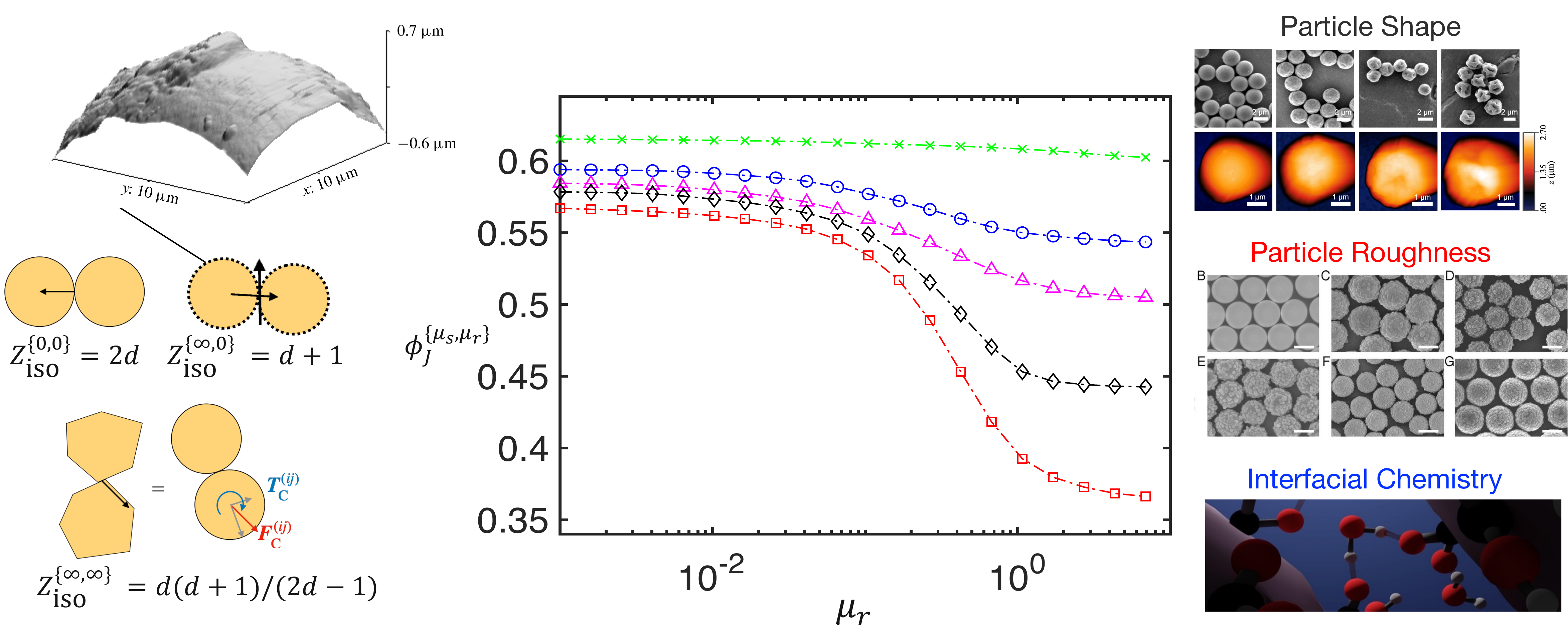}	
	\caption{\textbf{Connecting Jamming and Particle Features}. \emph{Left: Different types of constraints on relative particle motion}: Hard sphere, frictionless $\{\mu_s,\mu_r\} = \{0,0\}$ having isostatic condition $Z_{\mathrm{iso}}^{\{0,0\}}=2d$ in $d-$ dimension leading to $\phi_J^{\{0,0\}} \approx 0.65$ in 3-$d$; infinite sliding friction $\{\mu_s,\mu_r\} = \{\infty,0\}$ having isostatic condition $Z_{\mathrm{iso}}^{\{\infty,0\}}=d+1$ in $d-$ dimension leading to $\phi_J^{\{0,0\}} \approx 0.57$ in 3-$d$; infinite sliding and rolling frictions $\{\mu_s,\mu_r\} = \{\infty,\infty\}$ having isostatic condition $Z_{\mathrm{iso}}^{\{\infty,\infty\}}=d(d+1)/(2d-1)$ in $d-$ dimension leading to $\phi_J^{\{0,0\}} \approx 0.365$ in 3-$d$. \emph{Middle: Jamming phase diagram:} Jamming volume fraction $\phi_J^{\{\mu_s,\mu_r\}}$ \textcolor{black}{extracted by fitting $\eta_r$ in the lubricated and frictional states to the functional form $\eta_r = (1-\phi/\phi_J^{\{\mu_s,\mu_r\}})^{-2}$} plotted as a function of $\mu_r$ for several values of $\mu_s$.  \emph{Right: Control knob of jamming by tuning particle properties}: (top) particle shape, (middle) roughness, and (bottom) interfacial chemistry. The shading on each panel corresponds to the range of jamming volume fraction (middle panel). The middle panel is adapted from Ref.~\cite{Singh_2022}. 
    } 
	\label{fig_constraints}%
\end{figure*}

\section{Friction, constraints, and a minimal framework for the rheology of dense suspensions}
Once frictional contacts form between particles in dense suspensions, the central question shifts from how these interactions are modeled to how they reorganize stress transmission and affect flow, which we discuss below.
\paragraph{\textit{Constitutive expectations for dense suspensions}} 
%
Dense suspensions are commonly studied under simple steady shear, either at a fixed volume fraction $\phi$ or particle pressure $\Pi$~\cite{guazzelli_2018, Boyer_2011}. In a volume-controlled geometry, steady-state stresses take the form
\begin{equation}
\sigma_{xy} = \eta_0 \dot{\gamma}\,\eta_s(\phi), \qquad
\Pi = \eta_0 \dot{\gamma}\,\eta_n(\phi),
\end{equation}
where $\eta_0$ is the suspending fluid viscosity and $\eta_s$ and $\eta_n$ are the shear and
normal viscosities that diverge as $\phi$ approaches the jamming volume fraction $\phi_J$~\cite{Boyer_2011}.
%

In a pressure-controlled geometry, rheology is conveniently expressed in terms of the viscous number $J \equiv \eta_0 \dot{\gamma}/\Pi$, as follows:
\begin{equation}
\mu = \mu(J), \qquad \phi = \phi(J),
\end{equation}
with $\phi \to \phi_J$ and $\mu \to \mu_c$ as $J \to 0$~\cite{Boyer_2011, Wang_2015}. Here, the macroscopic friction coefficient $\mu \equiv \sigma_{xy}/\Pi$ and volume fraction $\phi$ are the order parameters.
These constitutive forms closely parallel dry granular rheology, highlighting the central role of particle contacts in dense suspensions~\cite{Boyer_2011, clavaud2025quick}. However, it does not predict any rate dependence, which is striking: simulations with hydrodynamic-only and those that combine hydrodynamics with contact friction both produce steady states that are invariant under rescaling of the imposed shear rate~\cite{clavaud2025quick}. This is \emph{counterintuitive}, as particles of different shapes, sizes, and chemistries dispersed in different fluids exhibit DST (see the review ~\cite{Morris_2020} and the references therein).

\paragraph{\textit{Introducing a force scale: stress-activated frictional contacts}}
Hydrodynamics and friction alone are insufficient to produce a rate-dependent steady-state rheology. Thus, to obtain shear thickening, an additional physical ingredient, a force scale, must be introduced to establish the transition from lubricated (hydrodynamic) interactions to direct frictional contact.

In colloidal suspensions, Brownian motion provides this force scale. The colloidal force $F_B \sim k_B T / a$, competes with the imposed flow: at low $Pe$, the colloidal forces maintain particle separation and the suspension remains in a lubricated state; at high $Pe$, the particle contacts become frictional~\cite{Mari2015discontinuous, Morris_2020, Jamali_2019}.

However, for non-colloidal suspensions, no intrinsic force scale exists. This implies that shear thickening  arises from forces that \emph{protect} the particle surfaces at low stress. It can be numerically implemented in two ways: (i) the electrostatic repulsion model (ERM) and (ii) the critical load model (CLM). 

\textit{Electrostatic repulsion model (ERM)}: The electrostatic repulsion model originates from electrostatic double layers, steric stabilization, or other interfacial chemistry, and is often used for non-colloidal suspensions to prevent aggregation~\cite{Israelachvili_2011}. For two particles or radii $a_i$ and $a_j$, it is modeled as 
\begin{equation}
    |\vec{F}^{i,j}_{R}| = \left\{
    \begin{array}{lr}
        F_0a_ia_j/(a_i+a_j)\exp(-\kappa h)\boldsymbol{n}_{ij} & \mathrm{if}~ h^{(i,j)} \geq 0\\
        F_0a_ia_j/(a_i+a_j)\boldsymbol{n}_{ij} & :\text{otherwise}  
    \end{array}
\right\}~,
\end{equation}
where $\kappa^{-1}$ is the Debye length. 

\textit{Critical load model (CLM)}: The so-called critical load model (CLM) is a minimal model that provides additional scaling for stress-dependence. In this model, a critical normal force $F_0$ is required to activate the frictional contact between two particles. When the normal force $F_n$ is smaller than the $F_0$, the particle interactions are frictionless, beyond which finite friction $\mu$ becomes active.
\begin{equation}
    |\vec{F}^{i,j}_{C,tan}| \leq \left\{
    \begin{array}{lr}
        \mu(|\vec{F}^{i,j}_{C,norm}|-F_{CL}) & : |\vec{F}^{i,j}_{C,norm}| \geq F_{CL}\\
        0 & :\text{otherwise}  
    \end{array}
\right\}
\end{equation}

Both models introduce a characteristic force scale $F_0$, which sets a corresponding stress scale $\sigma_0 \sim F_0 / a^2$. The particles remain separated by lubrication films when $\sigma \ll \sigma_0$, i.e., the rheology is frictionless. For $\sigma/\sigma_0 \gg 1$, almost all contacts are frictional, thus, the rheology is frictional. The two limits $\sigma/\sigma_0 \ll 1$ and $\sigma/\sigma_0 \gg 1$ are still quasi-Newtonian: the flow is rate-independent with finite normal stress differences~\cite{clavaud2025quick}.

\paragraph{\textit{From friction to constraints--a minimal framework for dense suspension rheology}} Once stress-activated frictional contacts are allowed, the central question becomes how these \emph{nanoscopic} interactions or particle scale features, such as roughness, shape, and surface chemistry, give rise to \emph{macroscopic} rheology. A useful and unifying perspective is to view friction as \emph{constraints} on relative particle motion~\cite{ Singh_2020, Santos_2020, Singh_2022} that restrict specific degrees of freedom (DOF., thus reducing the available DOF.
Mechanical stability follows a balance between constraints and force and torque balance conditions. Maxwell--type counting arguments define the isostatic coordination number $Z_{\mathrm{iso}}$, which determines the jamming volume fraction for each type of constraint~\cite{Hecke_2009,Behringer_2018}.

In a $d-$ dimensional system, for frictionless particles, only normal forces are transmitted, and the torque balance is trivially satisfied, leading to $Z_{\mathrm{iso}} = 2d$ and $\phi_J^{(0,0)} \approx 0.64$ in $3d$. Constraining the sliding motion reduces the isostatic condition to  $Z_{\mathrm{iso}} = d+1$, shifting to $\phi_J^{(\mu_s,\mu_r=0)} \approx 0.57$ in 3$d$~\cite{Singh_2018, Singh_2020, Mari_2014, Ness_2016}.
However, an additional rolling constraint further reduces $Z_{\mathrm{iso}} = d(d+1)/(2d-1)$ and $\phi_J^{(\mu_s,\mu_r=0)} \approx 0.365$ in 3$d$~\cite{Singh_2020}. Dense suspensions~\cite{Singh_2020} or, in general, particulate system~\cite{Santos_2020} jamming can be understood in terms of constraint-dependent $\phi_J^{\{\mu_s,\mu_r\}}$, the premise of the approach being that the loss in DOFs is more important than the exact physics/chemistry that leads to it.

\textcolor{black}{The connection between jamming and particle properties is summarized in Fig.~\ref{fig_constraints}. In this framework, the jamming point $\phi_J^{\{\mu_s,\mu_r\}}$ is extracted by fitting the steady-state viscosity in the lubricated and frictional states to the functional form $\eta_r = (1-\phi/\phi_J^{\{\mu_s,\mu_r\}})^{-\alpha}$, where most studies assume $\alpha \sim 2$~\cite{Guy_2015, Royer_2016, Singh_2018}; however, exponents are also reported to depend on particle properties~\cite{Mari_2014}.}



%

\paragraph{\textit{Stress-activated constraints and shear thickening}} The stress-dependent activation of constraints naturally leads to shear thickening. As the shear stress increases beyond $\sigma_0$, the constraints are activated, reducing the effective jamming volume fraction $\phi_J^{\{\mu_s,\mu_r\}}$ at a fixed $\phi$. 
The stress-activated constraint picture provides a minimal constitutive framework for shear thickening in dense suspensions. In the Wyart--Cates framework~\cite{Wyart_2014}, the fraction of frictional contacts $f(\sigma)$ is a \emph{stress-activated} order parameter.  Based on this, the effective jamming volume fraction can be expressed as
\begin{equation}
\phi_m(\sigma) = \phi_J^{(\mu_s,\mu_r)} f(\sigma) +
\phi_J^{(0,0)} \left[1 - f(\sigma)\right],
\end{equation}
where $\phi_J^{(0,0)}$ and $\phi_J^{(\mu_s,\mu_r)}$ are the jamming volume fractions for the unconstrained and constrained states, respectively.

The common functional form for $f(\sigma)$ is:
\begin{equation}
f(\sigma) = \exp\left(-\sigma_0/\sigma\right).
\end{equation}
Thus, the relative viscosity can be expressed as the distance from the stress-dependent jamming point as follows:
\begin{equation}
\eta_r \sim \left(1 - \phi / \phi_m(\sigma)\right)^{-2}~.
\end{equation}
Similar expressions can also be written for the normal stresses and particle pressure~\cite{Singh_2018, More_2020b}.  This framework unifies the CST, DST, and SJ in terms of stress-activated friction/constraints. CST is observed for $\phi \ll \phi_J^{(\mu_s,\mu_r)}$, which becomes stronger as $\phi$ increases. At a critical volume fraction $\phi_c \lesssim \phi_J^{(\mu_s,\mu_r)}$, DST is observed, whereas SJ occurs at high stresses for $\phi > \phi_J^{(\mu_s,\mu_r)}$.

\paragraph{\textit{Beyond stress-activated friction -- realistic steady-state rheology}}
Experimentally observed flow curves often exhibit shear thinning and yield stresses preceding shear thickening, implying the need for additional interparticle interactions. 
In simulations, this is captured by the inclusion of short-range repulsive or attractive forces in the force balance, as follows:
\begin{equation}
\vec{0} =
\boldsymbol{F}_{\mathrm{H}}(\boldsymbol{X},\boldsymbol{U}) +
\boldsymbol{F}_{\mathrm{C}}(\boldsymbol{X}) +
\boldsymbol{F}_{\mathrm{cons}}(\boldsymbol{X}),
\label{eq:LFDEM_general}
\end{equation}
where $\boldsymbol{F}_{\mathrm{cons}} = F_R + F_A$, where $F_R$ and $F_A$ are the repulsive and attractive forces, respectively. The attractive force is often modeled as a van der Waals form $F_A=A\bar{a}/12(h^2+\Delta)$, where $\bar{a}$ is the harmonic mean radius, $A$ is the Hamaker constant and $\Delta$ is used to avoid the singularity (so the contacts are allowed)~\cite{Singh_2019, More_2021, Pednekar_2017, Rathee_2021,wang2023effect}. 
%
Attraction arising from van der Waals forces or depletion effects can lead to strong shear thinning and even a finite yield stress~\cite{Singh_2019, Pednekar_2017,Rathee_2021,wang2023effect}. 
In these cases, the suspension is an unyielded soft solid for $\sigma<\sigma_y$, and the shear thins from infinite viscosity and shear thickens at even higher stresses. At higher attraction strength can eventually obscure shear thickening behavior~\cite{Singh_2019, Pednekar_2017,Rathee_2021, Gopalakrishnan_2004}. 

\begin{figure}[!hbtp]
	\centering 
	\includegraphics[width=0.4\textwidth]{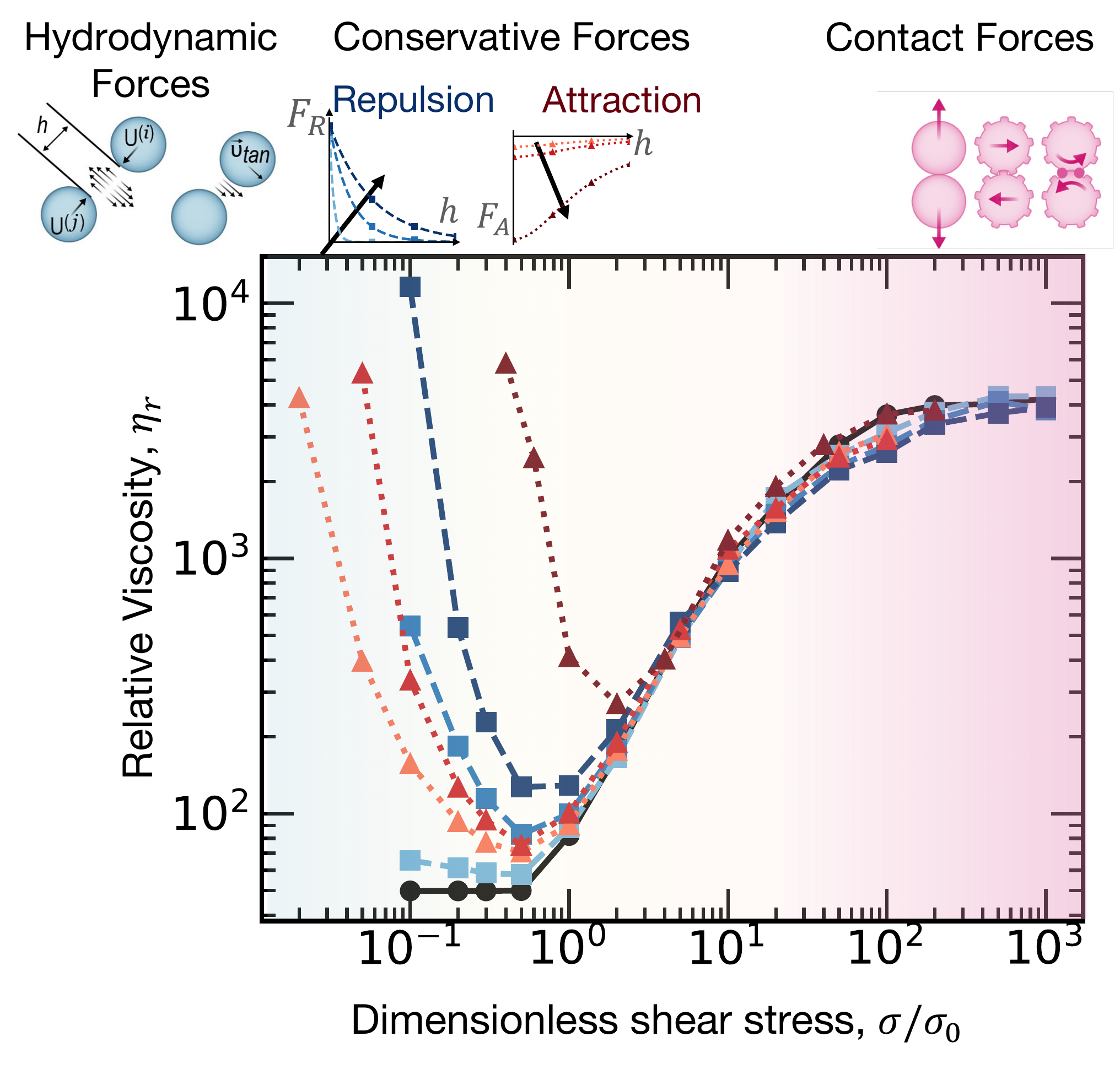}	
	\caption{\textbf{Connecting interparticle forces with rheology.} 
    \emph{Center: Representative shear-thickening response of a dense suspension},
showing relative viscosity $\eta_r$ as a function of dimensionless shear
stress $\sigma/\sigma_0$. The background shading indicates the dominant
force scale at a given stress: hydrodynamic and conservative interactions (low--intermediate stress) and frictional contact forces (high stress). Different symbols denote simulation models incorporating hydrodynamics and friction with varying conservative interactions. Circles (black solid line) correspond to the critical load model (CLM). Squares represent electrostatic repulsion, with darker shades indicating increasing Debye length $\kappa^{-1}$. Triangles denote simulations including both
repulsive and attractive forces, with darker shades corresponding to
a larger Hamaker constant $A$.
    } 
	\label{fig_rheo_forces}%
\end{figure}

Several mean-field models have been proposed to capture this physics: (i) the combination of Herschel--Bulkley with the WC model~\cite{Singh_2019}, and (ii) a more physics-based model of stress-released adhesive constraints~\cite{Guy_2018, Richards_2020}. The latter picture considers adhesive interactions that constrain relative rotational motion and are released as the stress increases. Thus, \emph{stress-released} constraints lead to yielding and shear-thinning, whereas stress-activated constraints lead to shear thickening. Adding the two together leads to the picture described above.

\paragraph{\textit{Role of inertia}} The discussion above assumes Stokes flow. When the particle inertia becomes finite, the rheology depends on the Stokes number $St$, and the system crosses over to an inertial regime \cite{Ness_2016,ge2020implementation,Trulsson_2012, Trulsson_2015}. 
%
A minimal representation is as follows: 
\begin{equation}
\Sigma \sim (\phi_J-\phi)^{\alpha}\dot{\gamma}^2
       + (\phi_J-\phi)^{\beta}\dot{\gamma}.
\end{equation}
\textcolor{black}{This framework attempts to integrate both viscous and inertial suspension physics, though it is important to note that the majority of experimental literature on shear thickening is in the viscous regime with $St, Re_p \ll 1$~\cite{Denn_2014, Mewis_2011}.}

\textcolor{black}{\paragraph{\textit{A note on model parameters}} 
Even though these simulation frameworks have semi-quantitatively reproduced experimentally observed steady-state rheology in the dense limit, several implementation choices require careful consideration:
\begin{itemize}
    \item \textbf{Near-hard-sphere limit:} 
As mentioned, a finite contact overlap (or compliance) is required to generate large-scale structures, i.e., perfectly rigid particles or the strict rigid limit cannot reproduce features such as DST, as shown by~\cite{townsend2017frictional}. Finite stiffness is essentially a numerical choice, with the maximum overlap decreasing with $k_n$. Previous simulation studies have shown that a maximum dimensionless overlap of approximately $2$--$3\%$ of the particle radius is generally sufficient to reproduce behavior consistent with experiments~\cite{Seto_2013a, Mari_2014, Singh_2018, Singh_2020}. Together with the normal stiffness, the tangential and rolling stiffnesses, $k_t$ and $k_r$, also require care to ensure that the spring stretches remain small compared to the particle size. Although no universal choice exists, commonly used ratios include $k_t/k_n = 2/7$~\cite{ge2020implementation} or $1/2$~\cite{Mari_2014, Singh_2018, Singh_2020}.
    \item \textbf{Lubrication regularization and friction:}  
Following the NHS idea, lubrication interactions need to be regularized, i.e., ($F_{\mathrm{lub}} \sim \frac{1}{h+\Delta}$). The choice of $\Delta$ requires care, as it effectively represents particle asperities or unresolved surface roughness, and typical values of $\Delta \sim 10^{-2} - 10^{-3}$ are commonly employed. The volume fractions for DST and SJ are directly related to the friction coefficients $\{\mu_s,\mu_r\}$. The families of friction coefficients $\{\mu_s,\mu_r\}=\{1,0\}, \{0.5,0.07\}$ have been shown to reproduce rheological trends that are semi-quantitatively consistent with experiments, although these friction values are somewhat larger than the experimentally measured ones.
    \item \textbf{Stokes number:} 
Stokesian Dynamics (SD)~\cite{Brady_1985} or related simulations integrating lubrication with contact friction simulate overdamped dynamics. These methods involve inversion of the hydrodynamic resistance matrix, which is numerically expensive. Other related methods that simulate small (but finite) Stokes numbers avoid explicit matrix inversion and instead use analytical expressions for pairwise lubrication interactions (see Refs.~\cite{Ness_2016, Ness_2023, monti2021fast, ge2020implementation, goyal2024flow}). Finally, these studies generally suggest that relatively small Stokes numbers, $(St \sim \mathcal{O}(10^{-2}))$, are sufficient for inertial effects to remain negligible.
%
\end{itemize}
}


\section{Recent Developments: Beyond Mean-field Description}
The mean-field models discussed so far, by construction, consider the globally averaged rheological description in terms of globally averaged order parameters, such as the mean frictional coordination number $Z$ or the fraction of frictional contacts $f(\sigma)$ (see review articles~\cite{morris2018lubricated, Morris_2020, Ness_2022, clavaud2025quick}). However, by definition, they miss the insight into stress heterogeneity, contact and force networks, and fluctuations in the rheological responses, often associated with features of a critical point in theoretical physics.





\paragraph{\textcolor{black}{\textit{Network science insights into dense suspension rheology:}}}
Insights from disordered particulate systems~\cite{Cates_1998a,Behringer_2018,hsiao2012role} have long suggested that mechanical stability and jamming are controlled by the emergence of system-spanning force chains and their organization under load. Here, we provide a brief overview of the recent progress along these lines.

\begin{figure*}
	\centering 
	\includegraphics[width=0.85\textwidth]{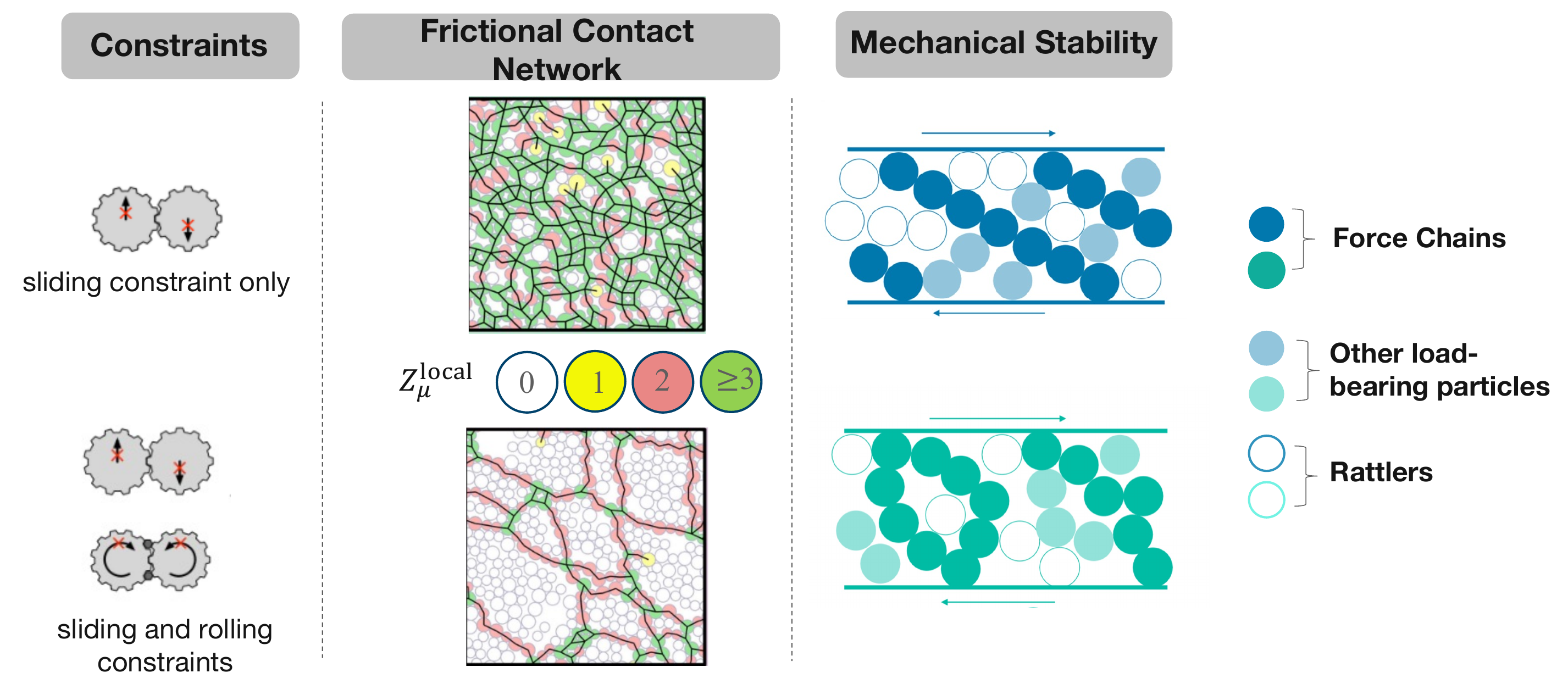}	
	\caption{\textbf{From microscopic constraints to mesoscale frictional contact network} \emph{Top: Sliding constraints only}: Considering sliding constraints only leads to a highly branched frictional contact network (FCN), with local frictional coordination number $Z_\mu^{\mathrm{local}} \ge 3$ (green colored) dominating, with a few particles with $Z_\mu^{\mathrm{local}} <2$ and minimal (\emph{yet finite}) rattlers ($Z_\mu^{\mathrm{local}} =0$). Strong force chains in the primary (compressive) axis need orthogonal support to maintain mechanical stability. 
    \emph{Bottom: Sliding and rolling constraints}: A suspension in which particles interact with both sliding and rolling constraints, similar rheology (viscosity $\eta_r$) has a very distinct FCN. The FCN is primarily composed of particles with ($Z_\mu^{\mathrm{local}} \le 2$) and a very few particles with ($Z_\mu^{\mathrm{local}} \ge 3$), and is composed of many rattlers. The mechanical stability for particles interacting with both sliding and rolling constraints does not require orthogonal support. Figure is adapted from Ref.~\cite{Sharma_2025}.
    } 
	\label{fig_mom0}%
\end{figure*}

This shifts the focus from traditional structural descriptors in real (grain) space, such as $g(r)$ or $S(q)$, to frictional contact or force network space~\cite{clavaud2025quick, Mari_2014}. At low stress, the frictional contact network (FCN) is not present. As $\sigma$ approaches $\sigma_0$, sparse frictional contacts are formed primarily along the compressive direction. At higher stress, contacts also develop in orthogonal directions, stabilizing force transmission and giving rise to looped and branched network structures that closely resemble the force chain organization in dry granular systems~\cite{Cates_1998a, Radjai_1998}.

Gameiro et al.~\cite{Gameiro_2020} showed that the persistence of looped structures in the FCN is correlated with the viscosity in two- and three-dimensional systems, over a range of packing fractions and applied stresses. Much of this understanding, particularly of network topology, has been developed using two-dimensional simulations. Under simple shear, structural variations are primarily confined to the flow and velocity-gradient directions, motivating the use of two-dimensional simulations. Consequently, much of the current understanding of frictional contact network topology has been developed within this framework.


From a network and graph-theoretic standpoint, particular attention has been given to loops as simple motifs in the FCN. Of particular importance are third-order loops, whose emergence has been linked to frustrated particle rotation and enhanced constraint~\cite{papadopoulos2018network}. Recent computational studies suggest that $\eta_r$ can collapse onto a master curve when plotted against the number of third-order loops $n_3$ across a wide parameter space of $(\phi,\sigma,\mu_s)$~\cite{d2025topological}. These observations suggest that FCN can serve as an important  \emph{mesoscale} descriptor of \emph{macroscale} rheology, including DST and SJ. Using the pebble-game algorithm, Naald et al.~\cite{Naald_2024} analyzed the locally ``minimally'' rigid clusters that are formed as the suspension is sheared. \emph{Nearly} system-spanning rigid clusters were found in the vicinity of jamming; however, the onset of rigidity $\phi_{\mathrm{rig}}$ was found to be close and yet distant from $\phi_{\mathrm{DST}}$ (packing fraction corresponding to DST).

%
Further studies suggest that the bulk rigidity stems from the local rigidity of FCNs~\cite{Santra_2025}. CST and DST have also been shown to be distinguished based on the degree of constraint and rigidity at the cluster level~\cite{Nabizadeh_2022}. Collectively, these results suggest that the emergence of rigidity may play an important role in the onset of shear jamming. However, the effect of many parameters, such as finite particle softness, system size, and protocol dependence, remains unexplored.


More recent studies have further generalized the network picture by explicitly accounting for different types of constraints, including only sliding or sliding and rolling~\cite{Sharma_2025}. Sliding-only systems exhibit highly branched networks that require orthogonal support for stability, as proposed by Cates et al.~\cite{Cates_1998a}. However, with rolling constraints, stability can be achieved through long self-supporting chains, even with a large fraction of rattlers. These results suggest that the network organization and mechanical stability depend not only on the number of contacts but also on the \emph{type} of constraints. These findings are consistent with experimental work by Hsiao and coworkers, who experimentally reported that the contact network is strongly affected by particle roughness~\cite{Pradeep_2021}.

In 3D, network-based measures of rigidity are less well established, and few structural descriptors have been investigated. Three-dimensional analysis of FCNs using $k$-core decompositions, in which $k$-core clusters are defined by particles with at least $k$ frictional contacts, suggests that simple contact percolation occurs in both CST and DST~\cite{Sedes_2022}.
Examining subnetworks of particles with fixed numbers of frictional contacts indicates that the percolation of particles with four frictional contacts coincides with the onset of the DST and is accompanied by large stress fluctuations that exhibit signatures of critical scaling~\cite{goyal2024flow}. 

Network-based analysis has emerged as a promising, still-evolving framework for understanding dense suspension rheology. These approaches suggest that rheological transitions, from discontinuous shear thickening to shear jamming, are closely linked to the reorganization of frictional contact networks, providing a mesoscale perspective that complements traditional constitutive descriptions.
\textcolor{black}{
It is now well established that the emergence of frictional contacts is closely related to continuous shear thickening (CST), while the formation of loop-like structures, corresponding to frictional contacts spanning both compressive and extensional directions, stabilizes the frictional contact network and enables load-bearing stress transmission associated with discontinuous shear thickening (DST) and shear jamming (SJ). Identifying universal topological descriptors governing rigidity, nonlocality, and shear jamming—particularly in three-dimensional systems—remains an active area of research.}


%


\paragraph{\textcolor{black}{\textit{Emergence of fluctuations and viewing DST/SJ as phase-transition like behavior}}} Discontinuous shear thickening (DST) has long been associated with the emergence of large fluctuations in both viscosity and normal stresses~\cite{Lootens_2003, Boersma_1991}. Experiments on highly concentrated, nearly monodisperse suspensions have shown that the viscosity can fluctuate by over an order of magnitude as the system approaches DST.
Motivated by the formation of frictional contacts, recent efforts have focused on understanding the microscopic origins of these fluctuations~\cite{Mari_2014, Heussinger_2013, goyal2024flow, Boromand_2018}. A common observation is that viscosity fluctuations are relatively small in both the lubricated and frictional regimes, where well-defined mean values exist (though they may differ significantly), whereas large temporal fluctuations emerge near the onset of DST, where the system intermittently samples both states.
This behavior can also be interpreted in terms of viscosity distributions: unimodal in the quasi-Newtonian regimes and becoming bimodal near the onset of DST. These observations have motivated connections to ideas from critical phenomena, where mean-field descriptions become insufficient and fluctuation-dominated behavior emerges.
More recent studies have explored scaling descriptions inspired by critical phenomena, suggesting that shear thickening may be associated with distinct underlying regimes and crossover behavior that can be characterized using scaling frameworks~\cite{Ramaswamy_2023, Malbranche_2023, ramaswamy2025universal}.

\section{Beyond Stokes flow, simple shear and viscometric flows}
Most simulation and theoretical descriptions of dense suspension rheology have focused on steady simple shear in periodic domains,  where the roles of hydrodynamic interactions, conservative forces, and frictional contacts can be isolated. However, real flows involve non-viscometric deformations, transients, confinement, and pressure-controlled conditions. A central question is therefore which physical mechanisms identified in simple shear persist under more general loading conditions.

\subsection{Extension of lubrication and friction approaches beyond Lees-Edwards imposed simple shear}
{\textit{Extensional flows:}} Particle-based simulations have extended dense-suspension modeling beyond simple shear to planar, uniaxial, and biaxial extension 
\cite{Cheal_2018, Seto_2017}. Frictionless suspensions exhibit nearly flow-type-independent jamming, consistent with approximately isotropic microstructure, whereas frictional systems show a pronounced dependence of the jamming volume fraction on the deformation geometry.

More recently, generalized periodic boundary conditions, such as the generalized Kraynik-Reinelt boundary conditions \cite{hunt2016periodic}, have been implemented to facilitate non-shear simulations of particulate materials, such as extensional flows, up to infinite strains. A recent study implemented these boundary conditions to simulate dense granular flows with varying deformation geometries ranging from triaxial extension to compression and pure shear, and a significant dependence of the stress response on the deformation geometry~\cite{clemmer2021shear}. The effects of the geometry of deformation are expected to be important in dense granular suspensions, suggesting the need for a complex tensorial rheological description, such as that formulated in a recent theoretical framework \cite{giusteri2018theoretical}.

\emph{Pressure-imposed flows.} Beyond volume-controlled shear flow simulations, pressure-controlled simulations allow the system to compact or dilate as necessary in response to externally imposed flow. Wall-bounded simulation frameworks with imposed particle pressure allow the solid volume fraction $\phi$ to adjust dynamically under shear~\cite{athani2025transients,Dong_2017}. These methods provide direct access to transient rheology and the coupled evolution of shear, particle pressure, and microstructure. They demonstrate that dense suspension flow is controlled by proximity to mechanical instability rather than by shear rate alone, reinforcing the connections with granular-flow physics.

 In addition to the imposed pressure, simulations of dense suspensions can also be conducted under applied shear stress rather than the customary applied shear rate. Previously, such stress-controlled simulations have successfully identified the physics of flow-arrest transition in colloidal suspensions \cite{Wang_2015} and dry granular materials \cite{srivastava2019flow,srivastava2022flow}. In stress-controlled simulations, the system under applied stress can flow or not, allowing precise quantification of the yield stress, the divergence of properties such as viscosity near the yield stress, and the stochasticity associated with the flow-arrest transition.


\emph{Wall-bound flows}: The presence of a wall modifies hydrodynamic interactions and stress transmission, since particle-wall lubrication forces and long-range wall-mediated flows alter microstructure and diffusion. Incorporating these effects into particle-resolved simulations presents significant computational challenges.
Swan et al. extended Stokesian Dynamics to account for confinement by constructing wall-corrected mobility formulations, demonstrating how particle–wall interactions influence rheology and short-time diffusion \cite{swan2010colloids, swan2011hydrodynamics}. Earlier pressure-driven simulations by Nott and Brady revealed shear-induced particle migration and accumulation in the center of the channels \cite{nott1994pressure}. More recent algorithmic developments have incorporated curved boundaries and wall-induced traction effects, enabling quantitative predictions of confinement-dependent transport and rheology \cite{Zia_2018}.
These studies collectively show that boundaries fundamentally alter microstructure, diffusion, and stress heterogeneity, and must be explicitly resolved to connect dense-suspension mechanics with realistic flow geometries.

\subsection{Grid-based Frameworks \label{OtherMethods}}
\textcolor{black}{This Perspective has so far focused primarily on Stokesian-Dynamics-like approaches for dense suspensions, particularly in the dense limit where near-contact hydrodynamics and frictional interactions dominate the rheology. In this section, we briefly discuss simulation methods beyond the Stokesian Dynamics family, many of which have traditionally been applied to dilute and semi-dilute suspensions, though recent developments have extended some of these approaches toward denser regimes. The methods discussed below are grid-based and offer additional flexibility in treating particle characteristics, such as shape and size distribution, boundary conditions (including confinement), and solvent rheology, ranging from Newtonian to non-Newtonian.
}


\emph{Force Coupling Method (FCM) and Fictitious Domain Method (FDM)}: Force Coupling Methods (FCM) represent particles as distributed force multipoles embedded within a fixed fluid grid, avoiding explicit inversion of the resistance-matrix while retaining hydrodynamic interactions~\cite{Maxey2010FCM}. Sub-grid lubrication corrections enable accurate microstructural and rheological predictions with favorable $\mathcal{O}(N \log N)$ scaling. \textcolor{black}{Force-coupling-based approaches have enabled simulations of non-inertial suspensions with $\mathcal{O}(10^3)$ particles while resolving many-body far-field hydrodynamic interactions~\cite{Abbas2007}, primarily in the dilute regime ($\phi \lesssim 0.2$). These studies also highlighted the importance of lubrication corrections for accurately capturing particle interactions at very small interparticle separations.
Addressing this, Dance and Maxey~\cite{Dance2003} incorporated lubrication interactions into the force-coupling method to study non-Brownian suspensions of inertialess particles in Stokesian Poiseuille flow. Using analytical near-field solutions, lubrication corrections were introduced for both particle--particle and particle--wall interactions, enabling more accurate treatment of near-contact hydrodynamics.}

Fictitious Domain Methods (FDM) enforce rigid-body motion through constraint-based formulations and fractional-step solvers without remeshing \cite{Gallier2014FDM, Gallier_2014}. Combined with frictional contacts using DEM and implicit time integration, these approaches capture the transition from hydrodynamic-dominated behavior in semi-dilute systems to contact-dominated rheology in the dense regime. Recent implementations within open-source frameworks have extended these methods to pressure-driven and heterogeneous flows, revealing non-local stress effects in confined geometries \cite{Orsi2023OpenFOAM}. \textcolor{black}{Building on such particle-resolved approaches, pressure-driven flow of dense non-Brownian suspensions with volume fractions up to $50\%$ was investigated in heterogeneous channel geometries~\cite{Orsi2024Mass}. These studies examined spatial variations in stress, volume fraction, and velocity profiles, revealing distinct wall-layering regions, a central plug-like region, and an intermediate bulk region. While the intermediate region showed good agreement with predictions from the local Suspension Balance Model (SBM), significant deviations were observed near the walls and in the central region, suggesting the importance of non-local stress contributions and highlighting the challenges of applying purely local constitutive descriptions to heterogeneous dense suspension flows.
Using both force-coupling and fictitious-domain methods, bidisperse non-Brownian suspensions under Couette flow were investigated to examine the role of particle-size variation and migration in heterogeneous flows~\cite{Howard2022}. Comparisons with a modified Suspension Balance Model (SBM) for bidisperse systems showed that the continuum description was capable of reproducing migration trends of the smaller particles in reasonable agreement with particle-resolved simulations. While the overall rheology exhibited relatively weak dependence on size bidispersity for the conditions studied, the work highlighted the growing need for continuum frameworks that incorporate particle-scale heterogeneity and migration dynamics.}


\emph{Immersed Boundary Method}: The Immersed Boundary Method (IBM) is well-suited for complex particle shapes and confined geometries, as both particles and boundaries are represented using discrete Lagrangian nodes. Combining the IBM with SD-type approaches enables the study of flexible fiber suspensions in Newtonian fluids, modeling fibers as inextensible slender bodies dynamically coupled to the surrounding flow~\cite{khan2023rheology}.

To address the challenges associated with property discontinuities in two-phase formulations, volume-filtered Navier–Stokes approaches have been developed \cite{vanWachem2024PCIBM}, introducing analytical viscous closures to maintain Galilean invariance and allowing coarse DNS in the dense limit. Fully particle-resolved DNS frameworks have also coupled IBM with Newton–Euler dynamics and DEM-based contact models to simulate dense suspensions in confined shear flows \cite{Breugem2024Friction}. These studies demonstrate that particle-resolved DNS can capture the coupled roles of hydrodynamics, lubrication, and contact interactions 
across various concentration regimes.

A limitation of classical SD-type methods is their restriction to Newtonian solvents. To overcome this, Shaqfeh and coworkers~\cite {Shaqfeh2023P1,Shaqfeh2023P2} developed particle-resolved DNS frameworks employing body-fitted and immersed-boundary grids to study suspensions in polymeric solvents modeled using Oldroyd-B and Giesekus constitutive equations. The evolution equations for the polymer stress tensor were coupled to the momentum balance via a source term in a finite-volume solver, enabling fully resolved two-way coupling between the particle motion and viscoelastic stresses. In the dilute-to-semidilute regime, these simulations showed that while the per-particle contribution may exhibit shear thickening, the overall suspension rheology can transition to shear thinning as the particle proximity suppresses particle-induced fluid stresses.\textcolor{black}{Using a Giesekus constitutive model, the effects of flow conditions, solvent rheology, and particle concentration on drag forces and normal stress differences can be systematically investigated for suspensions of fixed spherical particles~\cite{Ayar2023}.}

\begin{figure}[!htbp]
\centering\includegraphics[width=0.45\textwidth]{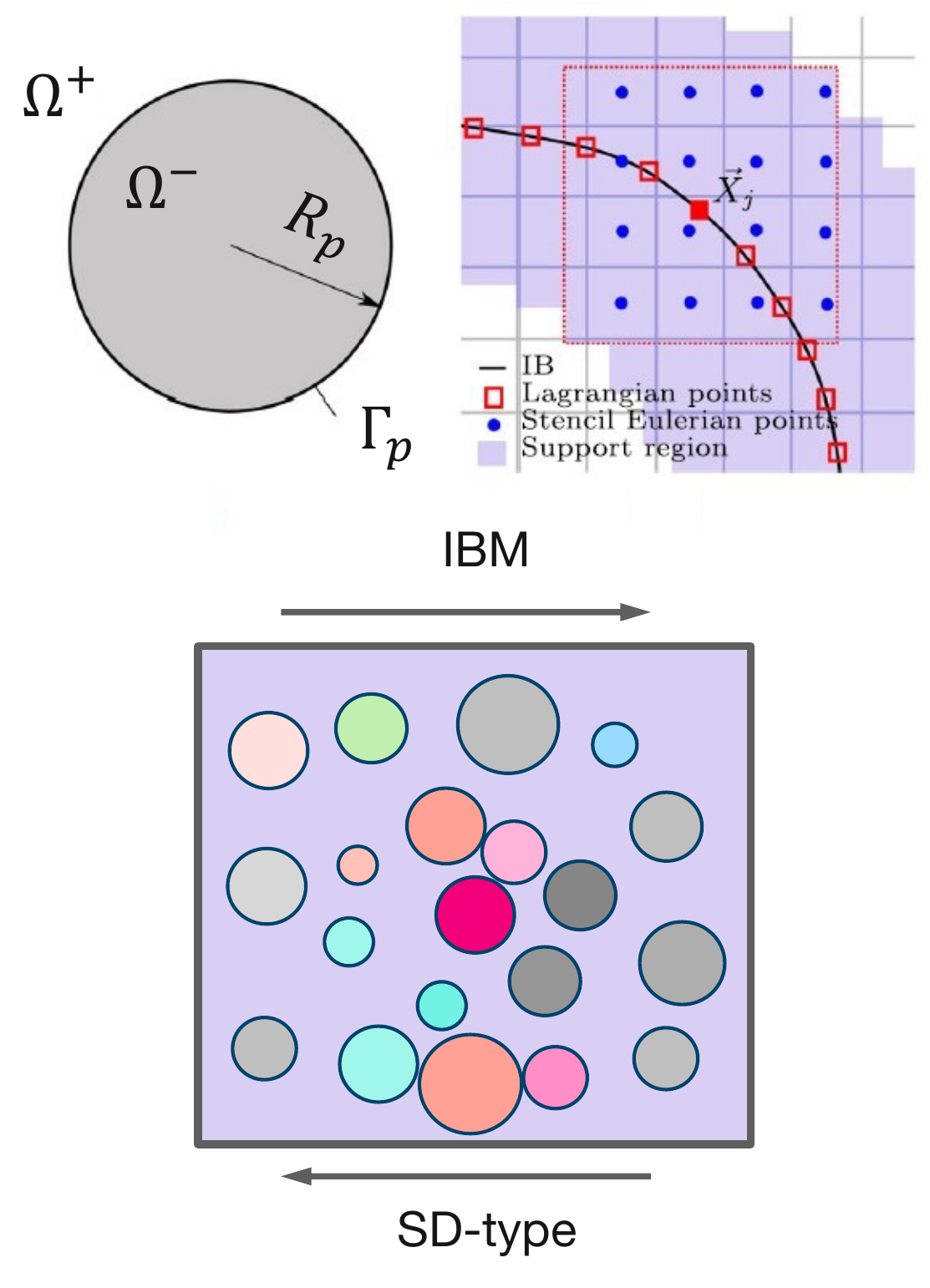}	\caption{\textbf{Spatial discretization of a suspended body in surrounding fluid} \emph{Top: Particle with radius $R_p$ and solid boundary $\Gamma_p$ `immersed` in a fluid: Solid boundary of the particle is marked by Lagrangian points to define it.}\emph{Bottom: Simple shear flow of polydisperse particles. Lagrangian points that define the particle boundaries interact with fluid, external Lagrangian points that belong to other suspended bodies and boundaries of the system.} Top panel is adapted from Reference~\cite{Azis2019}.}\label{fig_IBM}
\end{figure}

\subsection{Continuum Modeling} In engineering practice and natural phenomena, dense suspensions typically undergo a transient evolution through complex geometries. Predicting the fluid mechanics of dense suspensions in realistic flow scenarios requires a general, predictive $3$D continuum model that can be implemented and solved numerically using computational fluid dynamics (CFD). Most previous studies on dense suspensions primarily in the physics literature have focused on steady simple shearing flows~\cite{Seto_2013a, Singh_2018, Singh_2020, morris2018lubricated, Morris_2020}; however, these materials exhibit rich rheological features that need to be correctly accounted for in a constitutive model to accurately capture their dynamics in real applications. 

\textcolor{black}{
Continuum modeling of dense suspensions can be broadly classified into two categories. In a first class of continuum models, the two phases of the suspension, the particle phase and the fluid phase, are modeled individually as two interacting continua where each phase has its own continuum fields of velocity and volume fraction. Here, the particles are homogenized into a single continuum `granular' phase. In this formulation, a set of coupled partial differential equations (PDEs) can be written by considering basic conservation laws of mass, momentum, energy, and entropic inequality for each phase \cite{jackson2000dynamics,baumgarten2019general}. The resulting system of PDEs is generally not closed, and constitutive closures (models) that encode the physics of suspensions are required to model drag and buoyancy forces in the momentum exchange, fluid phase pore pressure, and stresses in the fluid and granular phases. Such ``two-fluid'' modeling is typically important in scenarios where the velocities of the two phases can be very different, such as in fluidized bed reactors \cite{agrawal2001role}. While some of the constitutive closures are well calibrated from experiments and simulations, the models for granular stresses represent key challenges in continuum modeling of dense suspensions. As discussed in detail above, this results from our recent understanding that in dense suspensions, frictional contacts strongly mediate suspension viscosity, particularly at very high solid volume fractions where CST, DST, and shear jamming are observed. In recent works \cite{baumgarten2019general,Baumgarten_2019}, the ``two-fluid'' representation was extended to model dense suspensions by expressing local suspension viscosity and granular pressure as a function of the fraction of frictional granular contacts \cite{Wyart_2014,Singh_2018}, which is included as an auxiliary field variable that locally modulates the volume fraction at which the suspension viscosity rapidly diverges, thus capturing the phenomena of CST, DST, and shear jamming.}

\textcolor{black}{
In a second class of continuum models, the suspension is modeled as a mixture, instead of two individual phases, through conservation PDEs of the mass, momentum and energy of the whole mixture. In these models, the migration or `drift' of particles relative to the mixtures requires additional closure modeling. For example, the diffusive flux of particles relative to the mixture is typically dependent on the spatial gradients of volume fraction and strain rate tensor \cite{mills1995rheology}, or can be expressed through a relationship with the divergence of the granular-phase pressure such as in the well-known ``suspension balance models'' \cite{nott1994pressure,nott2011suspension}. These types of models are usually valid when the drift velocities are small, which is often the case in steady flows of dense suspensions. Furthermore, higher-order rheological effects, such as normal stress differences have also been incorporated into such models of dense suspension in a frame-indifferent manner \cite{Morris_1999,badia2022frame}. These effects play a key role in secondary flow features such as curvature in free-surface flows, rod-climbing effect, and particle migration due to normal stress difference gradients \cite{guazzelli_2018}.}

\textcolor{black}{
The above continuum models do not explicitly account for the granular microstructure, and, as such, are well-suited for steady flows where the microstructure evolves smoothly and steadily. However, in highly transient flows such as shear reversal and oscillatory shear, an explicit accounting of the microstructure in the continuum modeling is essential to capture the complex interactions between the suspension rheology and microstructure \cite{Morris_2009}. In such cases, the microstructure is typically represented as a tensor, such as the fabric tensor, and its dynamic evolution is represented by a PDE that is coupled with the conservation PDEs \cite{gillissen2019constitutive,chacko2018shear}. The key challenge is such continuum models is an accurate calibration of the model parameters representing the microstructural evolution, which are typically obtained from extensive DEM simulations \cite{Seto_2017,Cheal_2018,srivastava2021viscometric}.}

\section{Take Home Messages and Outlook}
Dense suspensions are at the intersection of colloidal science, contact mechanics, continuum theory, and network physics, and remain central to industrial, technological, and geophysical processes. Over the past decade, major advances in particle-resolved simulations have transformed our understanding of the shear thickening, dilatancy, and shear jamming. Phenomena first reported nearly a century ago~\cite{Freundlich_1938} can now be reproduced with \emph{near-quantitative} accuracy, and their underlying mechanisms, hydrodynamic interactions, lubrication breakdown, frictional contacts, and mesoscale organization, can be systematically interrogated.

The computational advances discussed in this perspective have not only clarified the physics of stress-activated transitions but have also enabled the development of rational strategies to control and mitigate shear thickening, including orthogonal deformation protocols, particle deformation, and polymer-mediated interactions. Looking ahead, several challenges and opportunities remain.


\begin{enumerate}
    \item \emph{Going beyond nearly-similar-sized spherical particles}:  Most existing simulation frameworks focus on nearly equal-sized spherical particles, with limited treatment of strong polydispersity and non-spherical geometries \cite{Singh_2024, Malbranche_2023}. However, real materials exhibit broad size distributions and complex shapes. Extending computational methods to highly polydisperse and anisotropic particles is essential for industrial applications.
    \item \emph{Boundary Conditions and realistic geometries}:  Many state-of-the-art models assume idealized simple shear conditions without explicitly resolving wall slip, localization, free surfaces, or complex boundary conditions. Incorporating such boundary effects into particle-resolved and continuum descriptions is critical for theoretical rheological models and application-scale flows.
%
    \item \emph{Coupling different scales}: It is abundantly clear that the microscopic details of dense suspensions, such as particle characteristics and fluid rheology, strongly impact their macroscale behavior. Recent evidence also indicates that many intriguing physical phenomena in dense suspensions manifest at the mesoscale, such as the evolution of force chains and correlated particle motion. Therefore, multiscale modeling of these materials is required, enabling microscopic details to be reliably upscaled to macroscale continua while preserving the mesoscale characteristics of the system.
    \item \emph{Transients and dilatancy}: Although the physics of dense suspensions is well-established in the steady shear regime, practical applications of these flows routinely encounter more complex settings such as transient dilatancy, shear reversal, and non-shear deformation. Constitutive models capable of describing such complex flow histories with microstructural fidelity are still under active development.
    \item \emph{Beyond the mean-field description and statistical mechanics}: Increasing evidence indicates that dense suspensions cannot be fully characterized by scalar viscosity or mean-field closures alone. Incorporating network topology, force-chain statistics, and higher-order structural descriptors, particularly in three dimensions, offers a promising path toward predictive modeling near mechanical instability.
  \item \emph{Incorporating machine learning tools in modeling dense suspension mechanics:} Rapidly growing machine learning technologies provide exciting avenues to accelerate the simulations of dense suspensions, and consequently the discovery of physical mechanisms that underlie their mechanics~\cite{Jackson_2019,Barrat_2023}. For example, machine learning methods such as graph neural networks can be used to construct microstructural surrogates of the contact friction network, which has recently been shown to be promising in predicting the near-jamming physics of dense particulate systems ~\cite{aminimajd2025robust, aminimajd2025scalability,aminimajd2026graph, Mandal_2022}. Data-driven machine learning methods can be used to train complex rheological closures~\cite{Mangal_2025,brunton2020machine,lennon2023scientific} with appropriate physical constraints that upscale particle-level information into continuum descriptions capable of handling confinement, strong polydispersity,  and spatio-temporal stress heterogeneity. Such a training can also be done adaptively in-situ in a hybrid particle-continuum simulation with uncertainty quantification, as was successfully demonstrated recently for dense granular flows~\cite{siddani2025adaptive}. While the rapid increase in computational power and algorithmic advances will provide rich datasets of simulations of dense suspensions, crucial for the training of machine learning surrogates, the increasing availability of these surrogates will, in turn, accelerate the current capabilities of numerical simulations to length and time scales previously unachievable.
\end{enumerate}

\section*{Acknowledgements}
A.S. acknowledges financial support from the Case Western Reserve University startup funds. O.A. was funded by the American Chemical Society Petroleum Research Fund (Grant \# 68297-DNI9). I.S. and B.S. acknowledge support from the U.S. Department of Energy (DOE), Office of Science, Office of Advanced Scientific Computing Research, Applied Mathematics Program under Contract No. DE-AC02-05CH11231. This research was also supported in part by grant no. NSF PHY-2309135 to the Kavli Institute for Theoretical Physics (KITP).
\bibliographystyle{elsarticle-num} 
\bibliography{dst}
\end{document}